\documentclass[10pt]{blois07}
\usepackage{graphicx}
\usepackage{ccaption}
\captiondelim{. }

\def\gsim2{\scriptsize \mathrel{\rlap{\lower2pt\hbox{$\sim$}}\raise2pt\hbox{$>$}}}
\def\lsim2{\scriptsize \mathrel{\rlap{\lower2pt\hbox{$\sim$}}\raise2pt\hbox{$<$}}}
\begin{document}
\pagestyle{plain}
\title{\center{Deep-Elastic pp Scattering at LHC from Low-x Gluons}}
\author{\hspace*{4cm}M. M. Islam$^{a}$, J. Ka$\check{s}$par$^{b,c}$ and R. J. Luddy$^{d}$}
\institute{\center{
$^{a}$\footnotesize{Department of Physics, University of Connecticut, Storrs, CT 06269, USA, (islam@phys.uconn.edu)}\\ 
$^{b}$ TOTEM Collaboration, CERN, Geneva, Switzerland (Jan.Kaspar@cern.ch)\\
$^{c}$ Institute of Physics, Academy of Sciences of the Czech Republic, Prague\\
$^{d}$ Department of Physics, University of Connecticut, Storrs, CT 06269, USA, (RJLuddy@phys.uconn.edu)\\
}}

\maketitle{ }
\thispagestyle{plain}
\noindent{\bf{Abstract}}\rm
\vspace*{-0.2cm}
\begin{small}
\begin{flushleft}
Deep-elastic pp scattering at c.m. energy 14 TeV at LHC in the momentum transfer range 4 GeV$^{2}$ $\lsim2$ $\vert t \vert$ $\lsim2$ 10 GeV$^{2}$ is planned to be measured by the TOTEM group.  We study this process in a model where the deep-elastic scattering is due to a single hard collision of a valence quark from one proton with a valence quark from the other proton.  The hard collision originates from the low-x gluon cloud around one valence quark interacting with that of the other.  The low-x gluon cloud can be identified as color glass condensate and has size $\simeq$ 0.3 F.  Our prediction is that pp $\rm{d}\sigma/d\it{t}$ in the large $\vert t \vert$  region decreases smoothly as momentum transfer increases.  This is in contrast to the prediction of pp $\rm{d}\sigma/d\it{t}$ with visible oscillations and smaller cross sections by a large number of other models.
\end{flushleft}
\end{small}

\hspace*{-0.9cm}\it PACS:\rm 12.39.-x; 13.85.Dz; 14.20.Dh

\hspace*{-0.9cm}\it Keywords:\rm \small pp scattering; LHC/TOTEM; Proton structure; Low-x gluons; Color Glass Condensate \normalsize


   A major discovery from deep-inelastic ep scattering at HERA is the dramatic increase of low-x gluon density in a proton [1].  The HERA data have been well described by the saturation model of Golec-Biernat and W\"{u}sthoff (GBW model) [2,3].  The prediction of geometric scaling in this model for low-x region ($x<0.01$) has been impressively confirmed by Sta\'{s}to et al.[4].  In GBW model, the virtual photon splits into a dipole (a $\rm{q}\bar q$ pair) and the low-x gluons of a proton present an effective black disk to the dipole in the transverse $b$-plane.  In this paper, we present results of our investigation of deep-elastic pp scattering at LHC using key premises of GBW model pertaining to the low-x gluons.

      pp elastic scattering at LHC at center-of-mass energy 14 TeV and momentum transfer range $\vert t \vert \simeq$ 0 \textendash $ $ 10 GeV$^2$ is planned to be measured by the TOTEM group [5,6].  By deep-elastic pp scattering \textemdash $\;$we mean elastic $\rm{d}\sigma/d\it{t}$ distribution in the momentum transfer range 4 GeV$^2$ $\lsim2$ $\vert t \vert $ $\lsim2$ 10 GeV$^2$.  In this range, a proton probes the other proton at a transverse distance $b \sim q^{-1}$ ($q = \sqrt{\vert t \vert} $) and therefore at distances of one tenth of a fermi (F) or smaller.  High energy pp elastic scattering has been studied by us in a model of proton where the proton has an outer cloud of $\rm{q}\bar q$ condensed ground state (size $\simeq$ 0.86 F), an inner shell of baryonic charge density (size $\simeq$ 0.44 F) and an inner core of valence quarks (size $\simeq$ 0.2 F) (Fig. 1) [7].  Large $\vert t \vert $ pp elastic scattering in this model is due to a hard collision between a valence quark from one proton with a valence quark from the other proton \textemdash $\;$where the collision carries off the whole momentum transfer $\vert t \vert $ (Fig. 2).  In a previous paper, we have studied the case where this collision is due to a hard pomeron (a BFKL pomeron plus its next-to-leading order corrections) [8].  We now want to study the case where the low-x gluon cloud of one valence quark interacts with that of the other valence quark leading to the deep-elastic scattering.

      The valence quarks in our model are color neutral quasi-quarks, because of the gluon clouds around them.  To describe the valence quark-quark (q-q) scattering, we consider first an eikonal description for a purely absorptive process where the scattering amplitude is given by

$\hat{T}(\hat{s},t)=i\,\hat{p}\,\hat{W}\frac{1}{2\pi}\int{\rm{d}^{2}\it{b}\;e^{i \vec{b} \cdot \vec{q}}[1-\rm{exp}\it(-\hat{\chi}_{\rm{I}}(\hat{s},b))] }$;\hspace{5.1cm}               (1)\vspace{0.05cm}\\
$\hat{\chi}_{\rm{I}}(\hat{s},b)$ is the imaginary part of the eikonal and the hat (\textasciicircum ) refers to valence quark-quark scattering. From Eq.(1), we obtain the Born (single scattering) amplitude

$\hat{T}_{\rm{B}}(\hat{s},t)=i\,\hat{p}\,\hat{W}\;\frac{1}{2\pi}\int{\rm{d}^{2}\it{b}\;e^{i\; \vec{b} \cdot \vec{q}}\;\hat{\chi_{\rm{I}}}(\hat{s},b)}$. \hspace{6.62cm} (2)\\
Applying the optical theorem: $Im \;\hat{T}(\hat{s},0)=\frac{\hat{p}\;\hat{W}}{4\pi}\;\hat{\sigma}_{\rm{tot}}(\hat{s})$ to Eq. (1) and, in Born approximation, to Eq. (2), we find

$\hat{\sigma}_{\rm{tot}}(\hat{s})=2\,\int{\rm{d}^{2}\it{b}\;[1-\rm{exp}\it (-\hat{\chi}_{\rm{I}}(\hat{s},b))] }$\hspace{6.96cm}                 (3)\\
and

$\hat{\sigma}_{\rm{B}}(\hat{\it{s}})=2\,\int{\rm{d}^{2}\it{b}\;\hat{\chi}_{\rm{I}}(\hat{s},b)}$,\hspace{9.16cm}    (4)\\
where $\hat{\sigma}_{\rm{B}}(\hat{\it{s}})$ is the total cross section in Born approximation.  Let us next consider a factorizable eikonal: $\hat{\chi}_{\rm{I}}(\hat{s},b)=g(\hat{\it{s}})\;\frac{1}{2}\;F(b)$, where $g(\hat{s})$ is an unknown real function of $\hat{s}$ and $F(b)$ is a normalized $b$-profile: $\int{\rm{d}^{2}\it{b}\;F(b)}=1$.  Then Eq. (4) leads to

$\hat{\sigma}_{B}(\hat{s})=g(\hat{s})$, \hspace{10.9cm}     (5)\\
i.e. the unknown function $g(\hat{s})$ has to be the total cross section in Born approximation.  The corresponding total and Born 
amplitudes from Eqs. (1) and (2) are

$\hat{T}(\hat{s},t)=i\,\hat{p}\,\hat{W}\frac{1}{2\pi}\int{\rm{d}^{2}\it{b}\;e^{i \vec{b} \cdot \vec{q}}[1-\rm{exp}\it(-\hat{\sigma}_{\rm{B}}(\hat{s})\frac{1}{2}\;F(b))] }$ \hspace{4.3cm}   (6)\vspace{0.05cm}\\
and

$\hat{T}_{\rm{B}}(\hat{s},t)=i\,\hat{p}\,\hat{W}\hat{\sigma}_{\rm{B}}(\hat{s})\;\frac{1}{4\pi}\int{\rm{d}^{2}\it{b}\;e^{i\; \vec{b} \cdot \vec{q}}F(b)}$. \hspace{6.28cm} (7)\vspace{0.1cm}\\
Eqs. (6) and (7) then show that to obtain elastic valence quark-quark scattering amplitudes (multiple or single), we need two physical quantities:  $\hat{\sigma}_{B}(\hat{s})$ and the $b$-profile of valence q-q scattering.

We view each valence quark as having a low-x gluon cloud around it described by a gluon density $g(x, Q_{s}^{2}(x))$ and a transverse space profile $f(b)$. (Our choice of the scale $Q_{s}^{2}(x)$ in gluon density comes from the GBW model, as will be seen later.) By noting that $g(x, Q_{s}^{2}(x))\rm{d}\it{x}\;f(b)\rm{d}^{2}\it{b}$ is the number of low-x gluons in the gluon cloud of a valence quark lying between fractional momentum $x$ and $x+\rm{d}\it{x}$ and in the $b$-plane area $\rm{d}^{2}\it{b}$, and by introducing a low-x gluon-gluon production cross section, we can write down a simple expression for valence q-q total cross section (which we take as Born cross section):

$\hat{\sigma}_{\rm{B}}(\hat{\it{s}})=\int{\rm{d}^{2}\it{b}}\;\int_{0}^{x_{c}}\;\rm{d}\it{x}_{1}\;g(x_{1},Q_{\rm{s}}^{\rm{2}}(\it{x}_{1}))\,\int{\rm{d}^{2}\it{b}_{1}\;f(b_{1})\int_{0}^{x_{c}}\;\rm{d}\it{x}_{2}\;g(x_{2},Q_{\rm{s}}^{\rm{2}}(\it{x}_{2}))}$ \vspace{0.10cm}\\
\hspace*{3.0cm} $\times \int{\rm{d}^{2}\it{b}_{2}\;f(b_{2})\;\tilde{\sigma}(x_{1}\;x_{2}\;\hat{s})\,\delta^{(\rm{2})}(\vec{b}+\vec{b}_{1}- \vec{b}_{2})}$;\hspace{3.9cm}             \vspace{0.05cm}(8)\\
here $\tilde{\sigma}(x_{1}\;x_{2}\;\hat{s})$ is the low-x gluon-gluon production cross section at c.m. energy squared $x_1 x_2 \hat{s}$ 
($\hat{s}$ is the c.m. energy of the colliding valence quarks: $\hat{s}=(p+k)^{2}\simeq p_{+}k_{-}$; see Fig. 2).  We refer to gluons lying in the range $0 < x < x_{\rm{c}}$ ($x_{\rm{c}}=0.01$) as low-x gluons.  $\hat{\sigma}_{\rm{B}}(\hat{\it{s}})$ can now be written as

$\hat{\sigma}_{\rm{B}}\it (\hat{s})=\sigma_{\rm{gg}}\it (\hat{s})\;\int{\rm{d}^{2}\it{b}\;F(b)}$,\hspace{8.65cm}         (9)\\
where

$\sigma_{\rm{gg}}\it(\hat{s})=\int_{0}^{x_{c}}\rm{d}\it{x}_{1}\;g(x_{1},Q_{\rm{s}}^{\rm{2}}(\it{x}_{1}))\,\int_{0}^{x_{c}}\rm{d}\it{x}_{2}\;g(x_{2},Q_{\rm{s}}^{\rm{2}}(\it{x}_{2}))\; \tilde{\sigma}(x_{1}\;x_{2}\;\hat{s})$ \hspace{2.42cm}(10)\\
and

$F(b)=\int{\rm{d}^{2}\it{b}_{1}\;f(b_{1})}\int{\rm{d}^{2}\it{b}_{2}\;f(b_{2})\;\delta^{(\rm{2})}(\vec{b} +\vec{b}_{1}-\vec{b}_{2})}$.
\hspace{4.45cm}                              (11)\vspace{0.05cm}\\
Equation (10) provides a low-x gluon-gluon total cross section and Eq. (11) provides the corresponding $b$-profile $F(b)$.  Comparing Eq. (9) with Eq. (4), we notice that

$2\;\hat{\chi}_{\rm{I}}(\hat{\rm{s}},b)=\sigma_{\rm{gg}}(\hat{s})\;F(b)$.\hspace{9.08cm}	                     (12)\vspace{0.1cm}\\
Furthermore, the low-x gluon $b$-profile $f(b)$ of each valence quark represents probability density in the $b$-plane and therefore is normalized: $\int \rm{d}^{2}\it{b}\;f(b)=1$.  Equation (11) then leads to 
$\int \rm{d}^{2}\it{b}\;F(b)=1$ and from Eq. (9), we obtain 
$\hat{\sigma}_{\rm{B}}(\hat{\it{s}})=\sigma_{\rm{gg}}(\hat{\it{s}})$.  Finally, from Eqs. (6) and (7), we find the q-q scattering amplitude to be

$\hat{T}(\hat{s},t)=i\,\hat{p}\,\hat{W}\frac{1}{2\pi}\int{\rm{d}^{2}\it{b}\;e^{i\; \vec{b} \cdot \vec{q}}[1-\rm{exp}\it (-\frac{1}{2}\sigma_{\rm{gg}}(\hat{s})\;F(b))]}$ \hspace{3.88cm}    (13)\vspace{0.1cm}\\
and the Born amplitude to be

$\hat{T}_{\rm{B}}(\hat{s},t)=i\,\hat{p}\,\hat{W}\sigma_{\rm{gg}}(\hat{s})\;\frac{1}{4\pi}\int{\rm{d}^{2}\it{b}\;e^{i\; \vec{b} \cdot \vec{q}}F(b)}$. \hspace{6.00cm} (14)

At this point, let us recall that our goal is to study deep-elastic pp scattering due to a single hard collision of a valence quark from one proton with that of the other (Fig. 2), where the collision carries off all the momentum transfer and originates from the low-x gluon cloud of one valence quark interacting strongly with that of the other.  We, therefore, focus on evaluating the Born amplitude, Eq. (14).  The expression for $\sigma_{\rm{gg}}(\hat{s})$, Eq. (10), shows that to evaluate it, we need to know the $x$-dependence of the gluon density $g(x,Q^{2}_{\rm{s}}(x))$.  The expression for $F(b)$, Eq. (11), shows that for its evaluation we need to know the $b$-profile $f(b)$ of the low-x gluon cloud of a valence quark.

We now want to point out that the analyses of deep-inelastic ep scattering in the GBW model, in fact, provide precise information on the $x$-dependence of the gluon density $g(x,Q^{2}_{\rm{s}}(x))$ and also insight into the $b$-dependence of the profile $f(b)$.  To understand this connection of GBW model with our work, we note that in deep-inelastic ep scattering, the virtual photon with large $Q^2$ probes very small distances inside the proton, and therefore the low-x gluon density it finds is that due to the valence quarks - given our model of the proton structure where valence quarks form the core of the proton (Fig. 1).  Furthermore, since the valence quarks are quasi-quarks, and as such are quite independent of each other, the low-x gluon density $g_{\mbox{\tiny{P}}}(x,Q^{2}_{\rm{s}}(x))$ of the proton probed by the photon can be considered as the sum of the low-x gluon densities of the valence quarks, i.e.

$g_{\mbox{\tiny{P}}}(x,Q^{2}_{s}(x))=3\;g(x,Q^{2}_{s}(x))$.  \hspace{8.19cm}                   (15)

In the GBW model, the virtual photon converts into a small $\rm{q\bar{q}}$ dipole which traverses the low-x gluon cloud of the proton, and the dipole-proton total cross section in impact parameter representation is

$\sigma_{\rm{tot}}^{\rm{dipole}}(x,r^{2})=2\,\int{\rm{d}^{2}\it{b}\;[1-\rm{exp}\it (-r^{\rm{2}}\;\frac{\pi^{\rm{2}}}{3}\alpha_{s}\;x\;g_{\mbox{\tiny{P}}}(x,\mu^{\rm{2}})\;\frac{1}{2}\;f(b))]}$,\hspace{2.67cm}    (16)\vspace{0.10cm}\\
where $f(b)$ represents the gluon distribution in the $b$-plane ($\int{\rm{d}^{2}\it{b}\;f(b)}=1$).  GBW next introduce a saturation scale $Q^{2}_{s}(x)$ related with the gluon distribution:

$\frac{4}{\sigma_0} \frac{\pi^{2}}{3}\alpha_{s}\;x\;g_{\mbox{\tiny{P}}}(x,Q_{s}^{2}(x))=Q_{s}^{2}(x)$ \hspace{7.85cm}        		(17)\vspace{0.05cm}\\
and take

$Q_{s}^{2}(x)=Q_{c}^{2}(\frac{x_{c}}{x})^{\lambda}$,\hspace{10.1cm}    (18)\vspace{0.05cm}\\
where $x_{c}$ is the cut-off value of $x$ mentioned earlier.  GBW take $\lambda=0.29$ from their analyses of data.  Equations (17) and (18) show that as $x\rightarrow 0$, $x\;g_{\mbox{\tiny{P}}}(x,Q_{s}^{2}(x))$  increases rapidly as $x^{-\lambda}$.  ($Q_{c}^{2}x_{c}^{\lambda}$ in Eq. (18) is connected with $Q_{0}^{2}\;x_{0}^{\lambda}$ of GBW: $Q_{c}^{2}\;x_{c}^{\lambda}=Q_{0}^{2}\;x_{0}^{\lambda}$).  From Eqs. (15), (17) and (18), we find

$x\;g(x,Q^{2}_{s}(x))=C(\frac{x_{c}}{x})^{\lambda}$, \hspace{9.1cm}(19)\vspace{0.05cm}\\
where $C$ is a constant.  Inserting this in Eq. (10), we obtain

\vspace{0.1cm}$\sigma_{gg}(\hat{s})=C^{2}\;\int_{0}^{x_{c}}\;\frac{\rm{d}\it{x}_{1}}{x_{1}}(\frac{x_{c}}{x_{1}})^{\lambda}\,\int_{0}^{x_{c}}\;\frac{\rm{d}\it{x}_{2}}{x_{2}}(\frac{x_{c}}{x_{2}})^{\lambda}\,\tilde{\sigma}(x_{1}\;x_{2}\;\hat{s})$. \hspace{4.4cm}       (20)\vspace{0.15cm}\\
This equation shows that the dominant contribution on the right-hand side comes from the region $x_{1}\rightarrow 0$, $x_{2}\rightarrow 0$; that is, from the threshhold region of $\tilde{\sigma}(x_1 \;x_2\;\hat{s})$.  Equation (20) can be written as a single integral over the c.m. energy squared $\zeta$ of the gluon-gluon collision:

$\sigma_{gg}(\hat{s})=C^{2}\;\int_{0}^{x_{c}^{2}\hat{s}}\;\frac{\rm{d}\zeta}{\zeta}(\frac{x_{c}^{2}\hat{s}}{\zeta})^{\lambda}\,\tilde{\sigma}(\zeta)$. \hspace{7.58cm}     (21)\vspace{0.1cm}\\
As the dominant contribution to the integral comes from the threshhold region, we can approximate $\tilde{\sigma}(\zeta)\simeq \tilde{\sigma}(\zeta_{0})\; \theta(\zeta-\zeta_{0})$, where $\zeta_{0}$ is the threshold value and $\tilde{\sigma}(\zeta_{0})$ is the threshold gluon-gluon cross section.  The above equation then becomes

$\sigma_{gg}(\hat{s})=C^{2}\;\tilde{\sigma}(\zeta_{0}) \int_{\zeta_{0}}^{x_{c}^{2}\hat{s}}\;\frac{\rm{d}\zeta}{\zeta}(\frac{x_{c}^{2}\hat{s}}{\zeta})^{\lambda}\,\rm{ln} (\frac{x_{c}^{2}\hat{s}}{\zeta})$. \hspace{6.32cm}     (22)\vspace{0.15cm}\\
The integral on the right-hand side of Eq. (22) can be carried out and leads to the result $\sigma_{gg}(\hat{s})\sim \hat{s}^{\lambda}\;ln \hat{s}$ for large $\hat{s}$.  We note that, if the threshold gluon-gluon cross section is dominated by production of dijets, then $\tilde{\sigma}(\zeta_{0})=\tilde{\sigma}_{\rm{dijet}}(4 m_{j}^{2})$, where $m_{j}$ is the average mass of a single jet.

We next turn to the $b$-dependence of the GBW model.  Using Eq. (17) in Eq. (16), we obtain

$\sigma_{\rm{tot}}^{\rm{dipole}}(x,r^{\it{2}})=2\,\int{\rm{d}^{2}\it{b}\;[1-\rm{exp}\it (-r^{\rm{2}}\frac{\sigma_{0}}{4}\;Q_{\rm{s}}^{2}(x)\;\frac{1}{2}\;f(b))]}$.\hspace{3.75cm}    (23)\vspace{0.1cm}\\
In GBW model, the low-x gluon cloud of a proton is taken as a black disk: $f(b)=(\pi R^2)^{-1}\; \theta (R-b)$, where $R$ is the black disk radius.  Identifying $\sigma_0$ as the black disk cross section, $\sigma_0=2 \pi R^2$, Eq. (23) becomes

$\sigma_{\rm{tot}}^{\rm{dipole}}(x,r^{\it{2}})=2\,\int{\rm{d}^{2}\it{b}\;[1-\rm{exp}(-\frac{1}{4}\;r^{\rm{2}}\;Q^{2}_{s}(x)\;\theta(R-b))]}$

\vspace{0.07cm}$\hspace{2.2cm}=\sigma_0 [1-\rm{exp}(-\frac{1}{4}\;r^{\rm{2}}\;Q^{2}_{s}(x))]$,\hspace{6.2cm}    (24)\vspace{0.10cm}\\
which is the total dipole cross section formula in GBW model.  This result, however, is problematic.  When the dipole-proton c.m. energy squared goes to infinity (i.e., $\frac{1}{x}\rightarrow \infty$), Eq. (24) leads to $\sigma_{\rm{tot}}(x,r^2)\rightarrow \sigma_0$ (a constant).  This, however, does not meet the expected saturation of the total cross section: $\sigma^{\rm{dipole}}_{\rm{tot}}(x,r^2)\sim \rm{ln}^{2} \frac{1}{x}$ as $x\rightarrow 0$, based on the Froissart-Martin (FM) bound [9-11].  On the other hand, if $f(b)\sim e^{-mb}$ for large $b$, then the FM bound is saturated.  Furthermore, numerical investigation of the QCD nonlinear evolution equation by Gotsman et al. shows that $f(b)$ falls off as $e^{-mb}$ for large $b$, when they exclude the kinematical region of large dipoles [12].  To address this problem, we require the low-x gluons of a valence quark to have a $b$-plane distribution $f(b)$, such that: 1) $f(b)$ falls off as $e^{-mb}$ for large $b$, 2) has finite value at $b=0$, and 3) satisfies the normalization condition $\int{\rm{d}^{2}\it{b}\;f(b)}=1$.

Defining the Fourier transform of $f(b)$ by $\hat{f}(q)$:

$\hat{f}(q)=\int{\rm{d}^{2}\it{b}\;e^{i\; \vec{b} \cdot \vec{q}}f(b)}$, \hspace{9.12cm}                                       (25)\\
we obtain from Eq. (14)

$\hat{T}_{\rm{B}}(\hat{s},t)=i\,\hat{p}\,\hat{W}\;\sigma_{\rm{gg}}(\hat{s})\frac{1}{4\pi}\;\hat{f}^{2}(q)$. \hspace{7.77cm}(26)\vspace{0.15cm}\\
As we remarked earlier after Eq. (22), $\sigma_{gg}(\hat{s})\sim \hat{s}^{\lambda}\;ln \hat{s}$.  However, since the power $\lambda$ is not precisely known, we drop the $\rm{ln}\hat{s}$ term and write Eq. (26) as

$\hat{T}_{\rm{B}}(\hat{s},t)=i\;\hat{s}\;\gamma_{\rm{gg}}\;\hat{s}^{\lambda}\;\hat{f}^{2}(q)$, \hspace{8.59cm}     (27)\vspace{0.05cm}\\
where $\gamma_{\rm{gg}}$ is a real positive constant.  Requiring $\hat{T}_{\rm{B}}(\hat{s},t)$ to be crossing even, i.e. high energy elastic valence q-q and q-$\rm{\bar{q}}$ scattering to be the same, modifies Eq. (27) to

$\hat{T}_{\rm{B}}(\hat{s},t)=i\;\hat{s}\;\gamma_{\rm{gg}}\;(\hat{s}\;e^{-i\;\frac{\pi}{2}})^{\lambda}\;\hat{f}^{2}(q)$. \hspace{7.28cm} (28)\vspace{0.05cm}\\
We next choose a general form for the low-x gluon distribution in the $b$-plane:

\vspace{0.1cm}$f(b)=\frac{1}{2\;\pi}\frac{m^{2}}{\Gamma(\mu+1)}(\frac{m\;b}{2})^{\mu}\;K_{\mu}(m\;b)$,\hspace{0.5cm}($\mu>0$)\hspace{5.91cm}   (29)\vspace{0.15cm}\\
which satisfies the three conditions we require of $f(b)$.  Equation (28) then becomes

\vspace{0.05cm}$\hat{T}_{\rm{B}}(\hat{\it{s}},t)=i\;\hat{s}\;\gamma_{\rm{gg}}\;(\hat{s}\;e^{-i\;\frac{\pi}{2}})^{\lambda}\;\frac{1}{\left (1+\frac{q^{2}}{m^{2}}\right )^{2(\mu+1)}}$\hspace{1cm} for low-x gluons.\hspace{2.45cm}   (30)

      As we mentioned earlier, previously we investigated valence quark-quark scattering where the valence quarks interacted via the hard pomeron, i.e. BFKL pomeron (reggeized gluon ladders) plus its next to leading order corrections [8].  Because of significant leading order corrections, the power of $\hat{s}$: $\hat{s}^{\omega}$ with $\omega=\omega_{\rm{BFKL}}$ is substantially reduced to $\omega\simeq$ 0.13 - 0.18 as argued by Brodsky et al.[13].  The Born amplitude we used was

\vspace{0.05cm}$\hat{T}_{\rm{B}}(\hat{\it{s}},t)\simeq i\;\hat{s}\;\gamma_{\rm{hp}}\;(\hat{\it{s}}\;e^{-i\;\frac{\pi}{2}})^{\omega}\;\frac{1}{\left (1+\frac{q^2}{m_{\rm{h}}^{2}}\right )}$\hspace{1.0cm} for the hard pomeron\hspace{2.56cm}		                   (31)\\
($m_{\rm{h}}=\it{r}_{0}^{-1}$, $\gamma_{\rm{hp}}=\gamma_{\rm{qq}}/m_{\rm{h}}^2$ using earlier notation).  Comparing Eqs. (30) and (31), we notice that the structure of the two Born amplitudes are the same.  This enables us to carry out present quantitative calculations the same way as before [8].  pp elastic scattering amplitude originating from the valence q-q scattering amplitude (Eq. (30)) due to low-x gluons is

\vspace{0.1cm}$T_{\rm{qq}}(\it{s},t)=i\;s\;\gamma_{\rm{gg}}\;(s\;e^{-i\;\frac{\pi}{\rm{2}}})^{\lambda}\;\frac{\mathcal{F}^{\rm{2}}(q_{\perp},\lambda)}{\left (1+\frac{q^{\rm{2}}}{m^{\rm{2}}}\right )^{\rm{2}(\mu+1)}}$,\hspace{5.92cm}                                     (32)\vspace{0.05cm}\\ 
where $s=(P+K)^{2}$ is the square of the c.m. energy of pp scattering (Fig. 2) and $\mathcal{F}^{\rm{2}}(q_{\perp},\lambda)$ are structure factors that arise when the momentum distributions of the colliding valence quarks inside protons are folded in.  Physically, each structure factor stems from the confinement of valence quarks inside the proton.

      Our predicted pp $\rm{d}\sigma/d\it{t}$ at LHC at $\sqrt{s}$ = 14 TeV for the whole momentum transfer range $\vert t \vert$ = 0 \textendash $ $ 10 GeV$^2$ due to the combined diffraction, $\omega$-exchange, and single hard q-q scattering from low-x gluons (i.e. q-q scattering due to the Born amplitude: Eq. (30)) is shown by the solid line in Fig. 3.  Also shown in the figure are $\rm{d}\sigma/d\it{t}$ due to each of these three processes separately.  As before, the parameters of the model were required to provide good fits of $\rm{p}\bar{p}\;$ $\rm{d}\sigma/d\it{t}$ at $\sqrt{s}$ = 546 GeV, 630 GeV and 1.8 TeV as well as satisfactory descriptions of $\sigma_{\rm{tot}}(\it{s})$ and $\rho(s)$ as a function of $\sqrt{s}$.  In our present calculations, the parameters $m$ and $\mu$ of Eq. (21) have the values $m$ = 1.67 GeV, and $\mu$ = $\frac{1}{4}$, while the parameter $\lambda$ responsible for the sharp rise of low-x gluon density has the value $\lambda$ = 0.29 as given by GBW.

      High energy pp scattering originating from valence quark-quark scattering has also been discussed by Kovner and Wiedemann [14] in a model where the proton is taken as a loosely bound state of three valence quarks.  At some initial high energy, they view each valence quark as a black (i.e. totally absorbing) disk with a gray periphery.  The black disk originates from saturated gluon density in the central region of a valence quark, while the gray periphery is due to the long range Coulomb-like field of massless gluons.  In valence q-q scattering, the interaction between the two gray regions leads to a power like growth of the total cross section and in their model explains the pre-asymptotic behavior of $\sigma_{\rm{tot}}\simeq 21.70\; s^{0.0808}$ due to the soft pomeron [15].  With increasing $s$, the black central region of each valence quark grows in transverse size.  Comparing the valence q-q scattering in KW model with that of ours, we notice that in our model it is the interaction of the low-x gluons of a valence quark with those of the other in the $b$-plane (Eq. (8)), which leads to a power like growth of the Born cross section $\sigma_{\rm{gg}}(\hat{s})\sim \hat{s}^{\lambda}$;  here $\lambda$ measures the rapid growth of the gluon density: $x\;g(x,Q_{s}^{2}(x))\sim x^{-\lambda}$.  When unitarized using the eikonal description (Eq. (13)), the power growth of the cross section will be tamed at asymptotic $\hat{s}$ by the nonperturbative exponential fall-off of $F(b)\sim e^{-m\;b}$ and will lead to a black disk of radius $b_{\rm{BD}}\simeq (\lambda/m)\;\rm{ln}\;(\hat{\it{s}}/s_{0})$ and $\sigma_{\rm{tot}}(\hat{s})\simeq 2\;\pi\;[(\lambda/m)\;\rm{ln}\;(\hat{\it{s}}/s_{0})]^2$.  Furthermore, the black disk will have an outer gray boundary region where the profile function $\hat{\Gamma}(\hat{s},b)=1-\rm{exp}\it (-\hat{\chi}_{\rm{I}}(\hat{s},b))$ falls from $\hat{\Gamma}(\hat{s},b)=1\;$ ($b < b_{\rm{BD}}$) to $\hat{\Gamma}(\hat{s},b)=0\;$ ($b > b_{\rm{BD}}$).  As $\hat{s}$ increases, the black disk radius will grow in transverse size and the gray region will move outward.  Since $f(b)\;\rm{d}^{2}\it{b}$ is the probability of a gluon to be in an area $\rm{d}^{2}\it{b}$ at a distance $b$ from the center of the valence quark (Eq. (8)), we can calculate the average size of the low-x gluon cloud surrounding a valence quark in the transverse plane; namely, $<b^{2}>^{1/2}$ where $<b^{2}> = \int{b^{2}\;f(b)\;\rm{d}^{2}\it{b}}$.  Using the values of the parameters $m$ = 1.67 GeV and $\mu =\frac{1}{4}$ for $f(b)$ (Eq. (20)), we obtain $<b^{2}>^{1/2}=0.27$ F, somewhat larger than the 0.2 F size used by KW in their estimate of $\sigma_{\rm{tot}}$, but in agreement with the value 0.3 F suggested by Kopeliovich et al. [16].

      High energy pp, $\rm{p\bar{p}}$ elastic scattering have been quantitatively studied by Block et al. in a QCD-inspired phenomenological model [17].  In the c.m. system, a proton is viewed in their model as a disk in the $b$-plane with a gluon distribution and a valence quark distribution.  Both $b$-plane distributions are described by dipole form factors.  Block et al. envisage gluon-gluon, gluon-quark and quark-quark interactions in elastic scattering, and derive the full scattering amplitude using an eikonal description.  Their work has been followed up by Luna et al. [18] who introduce a dynamical gluon mass as an infrared mass scale.  Calculation of the Born cross section in these models is similar to ours (Eqs. (9) and (20)), since these models incorporate the same low-x behavior of the gluon density.  However, these models do not associate low-x gluons with the gluon cloud of a valence quark.  Instead, the low-x gluons are taken to belong to the whole proton.

      We note that the low-x gluons we are considering are specified at the saturation scale $Q_{\rm{s}}^{2}(\it{x})$ (Eq. (17)) where $Q_{\rm{s}}^{2}(\it{x})=Q_{\rm{c}}^{\rm{2}}(\it{x}_{\rm{c}}/x)^{\lambda}$ (Eq. (18)).  In the phase space plot: $\rm{ln}(1/\it{x})$ versus $\rm{ln} \it{Q}^{\rm{2}}$, gluons lying in the region bounded by the saturation line $\rm{ln} \it{Q}^{\rm{2}}=\rm{ln} \it{Q}_{\rm{s}}^{\rm{2}}(x)$ and the line $\rm{ln} \it{Q}^{\rm{2}}=\rm{ln} \it{Q}_{\rm{c}}^{\rm{2}}(x)$ (separating the nonperturbative regime from the perturbative regime) are referred to as forming Color Glass Condensate (CGC) [19].  The low-x gluons in our calculations, therefore, lie on the boundary of the CGC region.  The CGC region, however, can be extended further in $\rm{ln} \it{Q}^{\rm{2}}$ using geometric scaling; namely, $Q^{\rm{2}}\leq Q_{\rm{ge}}^{2}(\it{x})$, $Q_{\rm{ge}}^{2}(\it{x})=Q_{\rm{s}}^{\rm{2}}(\it{x})[Q_{\rm{s}}^{\rm{2}}(\it{x})/Q_{\rm{c}}^{\rm{2}}]^{\rm{0.34}}$ [19].  As a consequence, the low-x gluons we are considering actually lie well within the CGC region and the low-x gluon cloud around a valence quark can be identified as CGC.  Deep-elastic valence quark-quark scattering is then due to CGC around one valence quark strongly interacting with that of the other.

      Our predicted pp elastic $\rm{d}\sigma/d\it{t}$ at LHC at c.m. energy 14 TeV and $\vert t \vert$ = 0 \textendash $ $ 10 GeV$^2$ shown in Fig. 3 reflects the structure of the proton given in Fig. 1.  In the small $\vert t \vert$ region, the outer cloud of one proton interacts with that of the other giving rise to diffraction scattering.  In the intermediate $\vert t \vert$ region (1 GeV$^{2} \lsim2 \vert t \vert  \lsim2$ 4 GeV$^2$), one proton probes the baryonic charge density of the other proton via vector meson $\omega$ exchange.  In the large $\vert t \vert$ region (4 GeV$^{2} \lsim2 \vert t \vert \lsim2 $ 10 GeV$^2$), one proton scatters off the other proton via a single hard valence quark-quark collision (Fig. 2).  The valence quark-quark collision underlying Fig. 3 originates from the low-x gluon cloud of one valence quark strongly interacting with that of the other.  An alternative mechanism, studied previously, is the valence quark-quark collision due to the hard pomeron exchange[8].  Both hard q-q scattering processes predict smooth decrease of $\rm{d}\sigma/d\it{t}$ in the large $\vert t \vert$ region and differential cross sections which are comparable within an order of magnitude.

      We observe that the proton structure we have arrived at from our phenomenological investigation of high energy pp and p$\bar{\rm{p}}$ elastic scattering is a chiral bag embedded in a quark-antiquark condensed ground state.   This structure (Fig. 1) is described by a nonperturbative effective field theory model \textemdash $ $ a gauged Gell-Mann-Levy linear sigma-model with spontaneous breakdown of chiral symmetry and confined valence quarks [7].  In contrast, perturbative QCD cannot provide a physical structure of the proton, nor can it provide quark confinement.  Furthermore, the valence quarks of perturbative QCD are current quarks, whereas those of nonperturbative effective field theory models are constituent quarks.  What all this implies is that perturbative QCD cannot provide much physical insight of nonperturbative physics.  As an example, the small size of the chiral bag $R\simeq 0.2$ F in our case indicates that the low energy properties of the nucleon as predicted by this model will be essentially the same as those of the chiral topological soliton model, which corresponds to $R\rightarrow 0$, i.e. with no explicit quark degrees of freedom [20].  Both the chiral bag model and the chiral topological soliton model, for instance, will predict an axial charge distribution of size $<r_{A}^{2}>^{1/2}\simeq 0.62$ F close to the experimental value [21].  From perturbative QCD, which has explicit quark degrees of freedom, one could not have foreseen this nonperturbative result.

      Finally, we note that pp elastic $\rm{d}\sigma/d\it{t}$ calculations at $\sqrt{s}$ = 14 TeV have also been carried out for the whole $\vert t \vert$ region 0 \textendash $ $ 10 GeV$^2$ using Block et al. model [17] as well as other models [22-24].  All these models predict visible oscillations as well as much smaller cross sections than ours in the large $\vert t \vert$ region (Fig. 4).  Therefore, precise measurement of elastic $\rm{d}\sigma/d\it{t}$ at large $\vert t \vert$ by the TOTEM group will be able to distinguish between our model and the other models and shed light on the dynamics of deep-elastic pp scattering.

\vspace{0.5cm}\noindent{\bf{Acknowledgment}}\rm

One of us (M.M.I.) would like to thank his colleague Alex Kovner for discussions.\\

\vspace*{3cm}
\bf{\noindent{References}}\rm

\begin{small}
\noindent 1. R. D. Heuer, A. Wagner, CERN COURIER 48 (2008) 34.\\
2. K. Golec-Biernat, M. W\"{u}sthoff, Phys. Rev. D 59 (1998) 014017.\\
3. K. Golec-Biernat, M. W\"{u}sthoff, Phys. Rev. D 60 (1999) 114023.\\
4. A. M. Sta\'{s}to, K. Golec-Biernat, J. Kwieci\'{n}ski, Phys. Rev. Lett. 86 (2001) 596.\\
5. TOTEM: Technical Design Report, January 2004, CERN-LHCC-2004-002.\\
6. M. Deile (TOTEM Collaboration), Proceedings of the 12th Int. Conf. on Elastic and Diffractive		 Scattering (2007), edited by J. Bartels, K. Borras, M. Diehl, H. Jung (to be published).\\
7. M. M. Islam, R. J. Luddy, A. V. Prokudin, Int. J. of Mod. Phys. A 21 (2006) 1.\\
8. M. M. Islam, R. J. Luddy, A. V. Prokudin, Phys. Lett. B 605 (2005) 115.\\
9. M. Froissart, Phys. Rev. 123 (1961) 1053.\\
10. A. Martin, Nuovo Cimento A42 (1966) 930.\\
11. L. Lukaszuk, A. Martin, Nuovo Cimento A52 (1967) 122.\\
12. E. Gotsman, M. Kozlov, E. Levin, U. Maor, E. Naftali, Nucl. Phys. A 742 (2004) 55.\\
13. S. J. Brodsky, V. S. Fadin, V. T. Kim, L. N. Lipatov, G. B. Pivovarov, JETP Lett. 70 (1999) 155.\\
14. A. Kovner, U. A. Wiedemann, Phys. Rev. D 66 (2002) 034031.\\
15. A. Donnachie, P. V. Landshoff, Phys. Lett. B 437 (1998) 408.\\
16. B. Z. Kopeliovich, I. K. Potashnikova, B. Povh, E. Predazzi, Phys. Rev. D 63 (2001) 054001.\\
17. M. M. Block, E. M. Gregores, F. Halzen, G. Pancheri, Phys. Rev D 60 (1999) 054024.\\
18. E. G. S. Luna, A. F. Martini, M. J. Menon, A. Mihara, A. A. Natale, Phys. Rev. D 72 (2005) 034019.\\
19. For a recent review of the field, see J. Jalilian-Marian, Y. V. Kovchegov, Progress in Particle and Nucl.	 Phys. 56 (2006) 104.\\
20. A. Hosaka and H. Toki, Physics Reports 277 (1996) 65.\\
21. U.-G. Meissner, M. Kaiser, W. Weise, Nucl. Phys. A 466 (1987) 685.\\
22. C. Bourrely, J. Soffer, T. T. Wu, Eur. Phys. J. C 28 (2003) 97.\\
23. P. Desgrolard, M. Giffon, E. Martynov, E. Predazzi, Eur. Phys. J. C 16 (2000) 499.\\
24. V. Petrov, E. Predazzi, A. V. Prokudin, Eur. Phys. J. C 28 (2003) 525.\\
\end{small}
%
\vspace*{1.00cm}
\begin{figure}[htp]
  \center{\hspace*{-0.0cm}
  \includegraphics[height=2.8in]{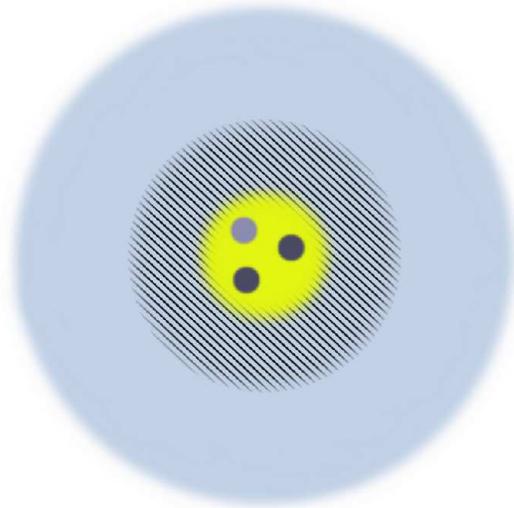}
\vspace*{-0.20cm}
\caption{
\small{
Model of proton structure underlying our study of high energy pp elastic scattering: proton has an outer cloud of $\rm{q}\bar q$ condensed ground state, an inner shell of baryonic charge density and an inner core of valence quarks.}
   }}
\end{figure}
\vspace*{2.5cm}
\begin{figure}[htp]
  \center{
  \includegraphics[height=1.6in]{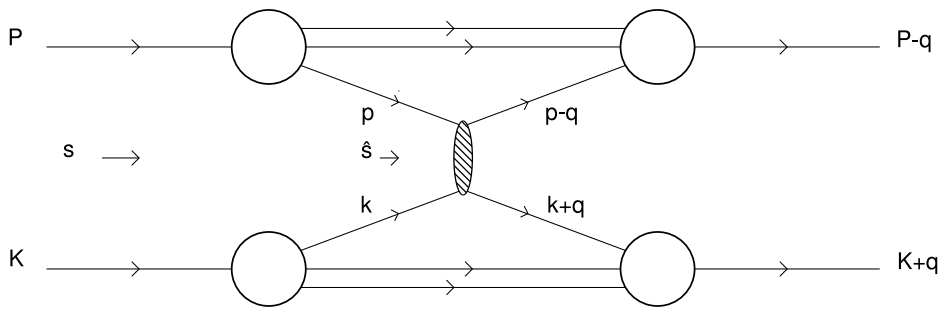}
  \vspace*{-0.2cm}
  \caption{
\small{Hard collision of a valence quark from one proton with a valence quark from the other proton.}
   }}
\end{figure}
\begin{figure}[htp]
  \vspace*{-1.00cm}
  \center{
  \includegraphics[height=3.6in]{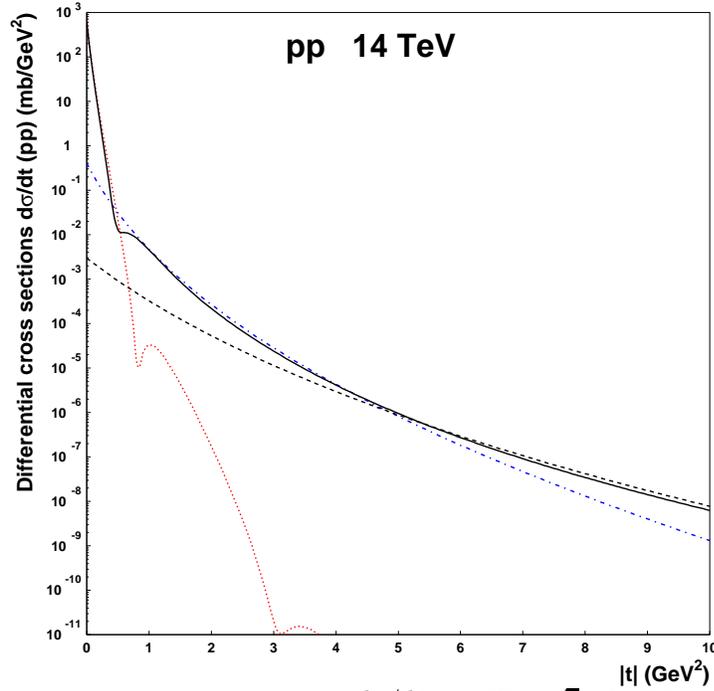}
  \vspace*{-0.5cm}
  \caption{
\small{Solid curve shows our predicted pp elastic $\rm{d}\sigma/d\it{t}$  at LHC at $\sqrt{s}$ = 14 TeV and $\vert t \vert$ = 0 \textendash $ $ 10 GeV$^2$ from combined diffraction, $\omega$ exchange and a single hard valence quark-quark collision due to their low-x gluon clouds.  Dotted curve shows $\rm{d}\sigma/d\it{t}$ due to diffraction only, dot-dashed curve that due to $\omega$ exchange only, and the dashed curve shows $\rm{d}\sigma/d\it{t}$ from the valence quark-quark collision only.}
   }}
\end{figure}
\begin{figure}[htp]
\center{
\vspace*{0.00cm}
\includegraphics[height=3.6in]{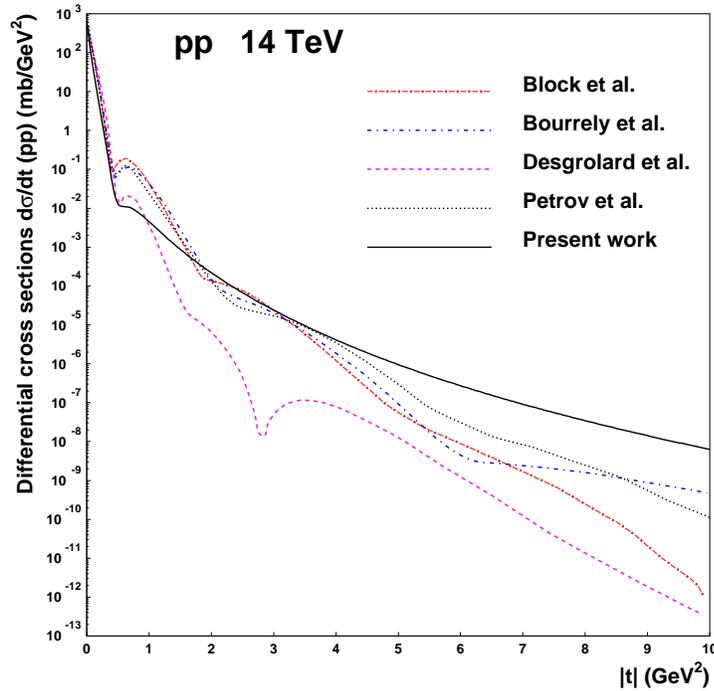}
\vspace{-0.4cm}
\caption{
\small{Comparison of our predicted $\rm{d}\sigma/d\it{t}$ (solid curve) at $\sqrt{s}$ = 14 TeV with the $\rm{d}\sigma/d\it{t}$ predicted by Block et al. [17], Bourrely et al. [20], Desgrolard et al. [21], and Petrov et al. (3 pomerons) [22] models.}
   }}
\end{figure}

\end{document}